\documentclass{anstrans}
\usepackage{graphicx} 
\usepackage{booktabs} 
\usepackage{microtype} 
\usepackage{color}
\usepackage[superscript]{cite}

\newcommand{\revision}[1]{{#1}}

\usepackage{layouts}
\usepackage{subfigure}
\usepackage[hidelinks]{hyperref}
\graphicspath{{Figures/}}

\usepackage{setspace}

\title{\MakeUppercase{A Thermal Discrete Element Analysis of EU Solid Breeder Blanket subjected to Neutron Irradiation} }

\author{Yixiang Gan,$^{a}$ Francisco Hernandez,$^{b}$ Dorian Hanaor,$^{a}$ Ratna Annabattula,$^{c}$ Marc Kamlah,$^{b}$ Pavel Pereslavtsev$^{b}$}

\institute{
$^{a}$ Particles and Grains Laboratory, 
School of Civil Engineering, 
The University of Sydney, 2006 NSW, Australia.
\and
$^{b}$ Karlsruhe Institute of Technology, 
D-76344, Eggenstein-Leopoldshafen, Germany.
\and
$^{c}$ Department of Mechanical Engineering,
Indian Institute of Technology - Madras, Chennai - 600036, India.
}

\begin{document}

\begin{abstract}
Due to neutron irradiation, solid breeder blankets are subjected to complex thermo-mechanical conditions. Within one breeder unit, the ceramic breeder bed is composed of spherical-shaped lithium orthosilicate pebbles, and as a type of granular material, it exhibits strong coupling between temperature and stress fields. In this paper, we study these thermo-mechanical problems by developing a thermal discrete element method (Thermal-DEM). This proposed simulation tool models each individual ceramic pebble as one element and considers grain-scale thermo-mechanical interactions between elements. A small section of solid breeder pebble bed in HCPB is modelled using thousands of individual pebbles and subjected to volumetric heating profiles calculated from neutronics under ITER-relevant conditions. We consider heat transfer at the grain-scale between pebbles through both solid-to-solid contacts and the interstitial gas phase, and we calculate stresses arising from thermal expansion of pebbles. The overall effective conductivity of the bed depends on the resulting compressive stress state during the neutronic heating. The thermal-DEM method proposed in this study provides the access to the grain-scale information, which is beneficial for HCPB design and breeder material optimization, and a better understanding of overall thermo-mechanical responses of the breeder units under fusion-relevant conditions. \footnote{\textit{Fusion Science and Technology}, \href{http://dx.doi.org/10.13182/FST13-727}{DOI: 10.13182/FST13-727}.}
\end{abstract}


\section{I. INTRODUCTION}
During operations of nuclear fusion breeder reactors, Helium-Cooled Pebble Beds (HCPB), as a candidate for fusion blankets \cite{Boccaccini2009}, are subjected to neutron fluxes to generate tritium to complete the fuel cycle. Thermo-mechanical conditions inside the breeder units are crucial for tritium generation and release rate \cite{Ying2012}, and the mechanical stability \cite{Gan2011} of the ceramic breeder materials, e.g., lithium orthosilicate, Li$_4$SiO$_4$ \cite{Knitter2007}. 
In order to gain an accurate understanding of the conditions during operations, thermo-mechanical models and experiments have been carried out for those materials \cite{Reimann2000, Reimann2002, Reimann2003, Dellorco2006, Zhao2013}, including  phenomenological and grain-scale approaches. 

Due to its granular nature, the solid breeder material exhibits strong thermo-mechanical coupling and heterogeneous behaviour. Experimental data \cite{Reimann2002, Reimann2003, ABOU-SENA2005} show a clear strain-dependent thermal conductivity, which raises questions concerning both the understanding of the underlying physics and the implementation of quantitative engineering analysis by considering this coupled thermo-mechanical behaviour. In phenomenological models, this coupling behaviour has been implemented based on empirical curves for effective thermal conductivity of ceramic beds from elaborated experimental measurements. The experimentally obtained effective thermal conductivity can be expressed as a function of both bed temperature ($T$) and strain ($\varepsilon$), i.e., $k_\mathrm{eff}(T,\varepsilon)$ \cite{Reimann2002, Reimann2003}. However, the empirical curves and coefficients may vary if different breeder materials, for instance lithium metatitanate, will be used or simply adopting different types of pebble size distributions of pebbles. Without accessing grain-scale information under operation-relevant conditions, the thermo-mechanical behaviour of the material can not be accurately assessed within breeder units.

Combining discrete and continuum approaches preserves both micro-scale material details and modelling capability at the system scale. Achieving this requires a multi-scale modelling scheme for bridging information at two different length scales. Statistical data can be achieved under given thermo-mechanical conditions, e.g., the force acting on individual pebbles \cite{An2007, gan2010,Annabattula2012},  and the crushing of pebbles \cite{Gan2011, Annabattula2012a, Zhao2013}.

To understand the thermo-mechanical coupling in the material with a granular form, we propose a numerical method, i.e., thermal discrete element method (Thermal-DEM \cite{gan2010, Gan2012}), to estimate the temperature and stress profiles within the ceramic breeder material under ITER-relevant conditions. Thermal-DEM models ceramic breeder pebbles as individual elements and describes the interactions between neighbouring pebbles, including not only contact forces but also heat fluxes, via solid-to-solid contacts and the interstitial gas phase between pebbles.
A typical EU design for breeder units \cite{Hernandez2013} and the relevant neutronic heating profile \cite{Pereslavtsev2010} have been adopted for the numerical analysis to demonstrate the feasibility of the proposed method for  the analysis of solid breeder blanket systems. 
By identifying key parameters that determine the resulting temperature profiles and stresses, this method can be incorporated for the design and optimisation of breeder materials in fusion blankets.

\section{II. METHOD}
\subsection{II.A. Thermal Discrete Element Method}

To study the heat transfer in pebble bed systems, we use the thermal discrete element method (Thermal-DEM), which is based on the conventional DEM \cite{Cundall1979} with one additional degree of freedom for the temperatures of the pebbles, to model individual pebbles and their thermo-mechanical interactions.

\begin{figure}[htb]
\begin{center}
\includegraphics[width=0.42\textwidth]{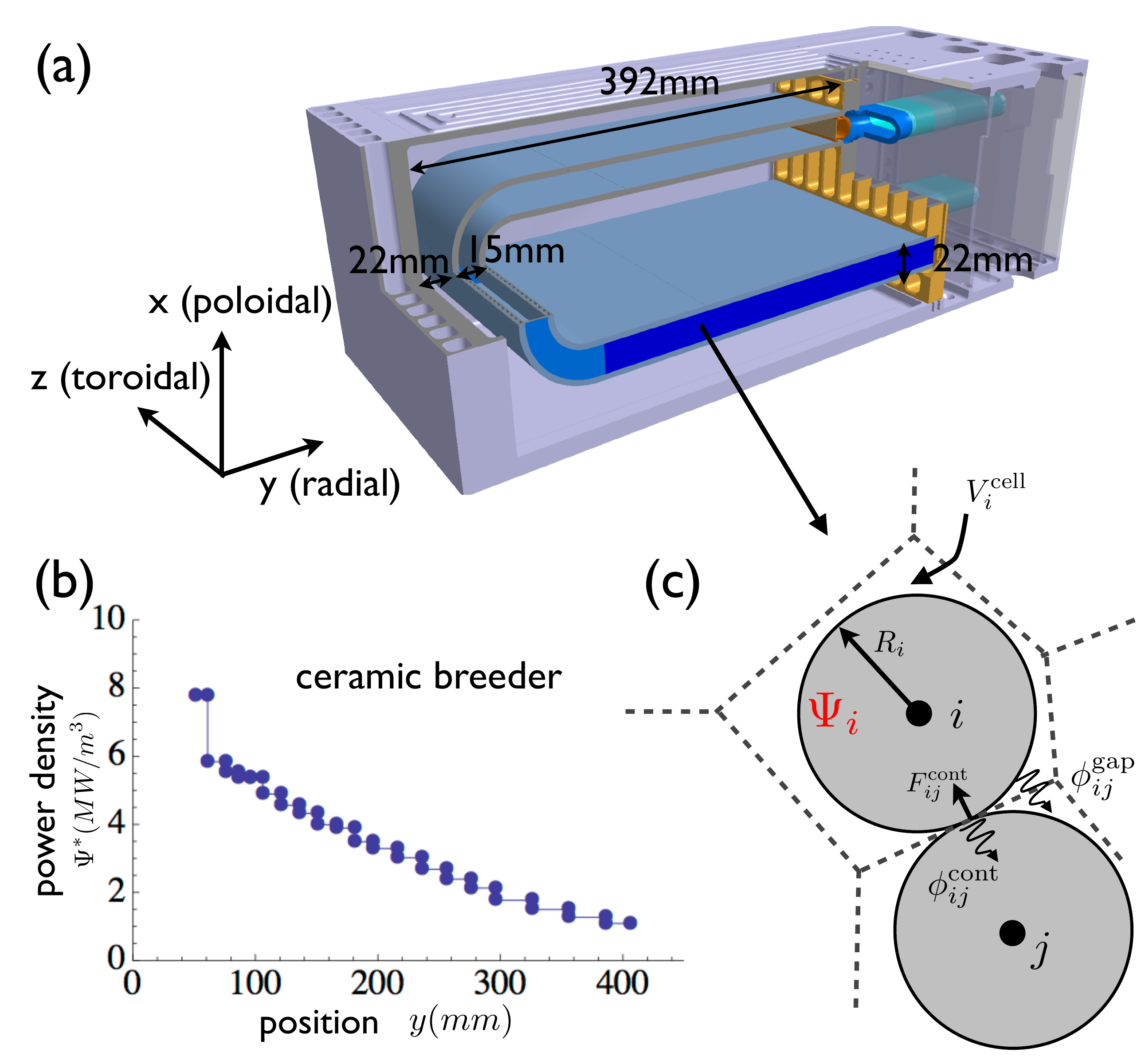}
\caption{Schematics of the approach: (a) EU design for ITER-TBM unit adopted from \cite{Hernandez2013}; (b) calculated power density of neutron irradiation during operation from \cite{Pereslavtsev2010}; (c) Thermal-DEM model for two pebbles, with mechanical interaction $F_{ij}$, volumetric heating $\Psi_i$, and heat fluxes $\phi_{ij}^\mathrm{cont}$ and $\phi_{ij}^\mathrm{gap}$ for solid-to-solid and gap conduction, respectively.}
\label{fig:DEM}
\end{center}
\end{figure}

The mechanical interactions between pebbles, which are modelled as elastic spheres, include the normal and frictional forces, based on the analytical solution for Hertzian contact. The positions of pebbles are updated according to Newton's law of motion with an explicit integration scheme, see more details in \cite{gan2010}.
For updating the temperatures of individual pebbles, the heat conduction between pebbles is described through Fourier's law. The thermal conductance between the pebbles consists of two parts: solid contact conductance, and the gap conductance through the interstitial gas phase. The inter-granular conduction is described by the analytical model proposed by \cite{Batchelar1977}, including the combined modes of heat transfer through both contact regions and gap regions. This analytical model considers the relative thermal conductivity between the solid and gas phases, as $k_s/k_g$, where $k_s$ denotes the conductivity of the bulk material and $k_g$ is the gas conductivity.
Moreover, the heat generation inside each pebble can be calculated by the power spectrum based on the neutron irradiation profile. This profile is introduced in Thermal-DEM as a source of volumetric heating.

Combining the heat transfer and heat generation by neutron irradiation, the rate of temperature change ($\dot T_i$) for the $i$-th pebble can be updated as
\begin{equation}
\dot T_i = \frac{1}{m_i c_p} (\sum_j \phi_{ij} + \Psi^* V_i^\mathrm{cell})
\mbox{ .}
\label{eq:tdot}
\end{equation}
Here, $m_i$ and $c_p$ are the mass of the pebble and the heat capacity of the bulk material, respectively. \revision{The heat flux $\phi_{ij}$ is calculated by}
\begin{equation}
 \phi_{ij} = - 2 \pi k_g (T_i-T_j) R_\mathrm{eff} \mathcal{H}(\alpha,\beta)
 \mbox{ .}
\end{equation}
\revision{The non-dimensional flux $\mathcal{H}$ is based on the relative thermal conductivity, $\alpha=k_s/k_g$, and relative contact area, $\beta=\alpha a/(2R_\mathrm{eff})$, where $a$ is the radius of contact area and $R_\mathrm{eff}=R_iR_j(R_i+R_j)^{-1}$ is the effective radius (more details can be found in \cite{Batchelar1977}).}

We adopt the volumetric heating profile from the neutronics calculation \cite{Pereslavtsev2010}, as $\Psi^*$ shown in Fig.\ \ref{fig:DEM}(b). The actual heating power in each pebble can be scaled based on its local volume and the power density, as $\Psi^* V_i^\mathrm{cell}$, where the cell volume, $V_i^\mathrm{cell}$, is obtained using a Voronoi tessellation for individual pebbles \cite{Rycroft2006}, shown schematically as the dashed cell in Fig.\ \ref{fig:DEM}(c). For each pebble, the scaled heating power density per solid volume is given by $\Psi_i = \Psi^* V_i^\mathrm{cell} / V_i^\mathrm{pebble}$, where $V_i^\mathrm{pebble}$ is the solid volume of the $i$-th pebble.

Thermo-mechanically coupled behaviours in the grain-scale model are automatically taken into account through the thermal expansion of pebbles. Under conditions of neutron heating, a compressive stress develops in the bed as the result of the pebbles' thermal expansion. In turn, the growing compressive stress enhances heat transfer through the enlargement of contact areas.

Based on the EU breeder unit design \cite{Hernandez2013} in Fig.\ \ref{fig:DEM}(a), the neutron heating profile for the ceramic breeder materials has been calculated in \cite{Pereslavtsev2010} for Li$_4$SiO$_4$. To demonstrate the feasibility of using this numerical method, three power densities have been selected, namely, 5, 7 and 8 MW/m$^3$, corresponding to different radial positions inside the breeder layer. The exact positions are not given in this study, since only the heating power density will be required from the blanket design.

\subsection{II.B. Material Parameters and Boundary Conditions}

The material properties of Li$_4$SiO$_4$ pebbles, compiled from different sources, are listed in Table \ref{tab:mat}. Those values have been calculated only at a fixed temperature of 500 $^\circ$C due to their weak temperature dependencies within the targeted temperature range. It would be straightforward to implement bulk material properties varying with temperature in the future work. 

The interstitial gas properties are listed in Table \ref{tab:wall}. Note that, relative to the bulk material, the thermal conductivity of the gas phase is sensitive to temperature. In this study, temperature-dependent gas conductivity has been implemented according to the empirical equation provided by \cite{Petersen1970}. The empirical equation used in this study includes also  pressure dependency, but the influence of gas pressure variation from 100 to 1000 kPa is negligible. A fixed purged gas pressure of 400 kPa is used in our simulations. It is noted that from experimental observations a higher interstitial gas pressure can lead to better bed conduction \cite{ABOU-SENA2005}. This is mainly due to the heat conduction via the gas phase in the regime of large Knudsen numbers ($K_n=\lambda/L$) depends on both the dimensions of the gaps ($L$) and the mean free path of the gas molecules ($\lambda$). In a granular media, the gap size ($L$) can be linked to the size of pebbles. This effect is not considered in this study, but it clearly will influence the effective conductivity of the beds, especially for beds with small pebbles.

\begin{table}[h]
\caption{Properties for Li$_4$SiO$_4$ at 500 $^\circ$C and porosity 5\%.}
\label{tab:mat}
\footnotesize
\centering
\begin{tabular}{l c l}
\hline \hline
Property & Symbol & Value \\
\hline
Young's modulus (GPa) \cite{FutamuraWeb} & $E$ & 92 \\
Poisson's ratio & $\nu$ & 0.24 \\
thermal conductivity (W/mK) \cite{ABOU-SENA2005} & $k_s$ & 2.743  \\
heat capacity (J/kg K) \cite{Billone1993} & $c_p$ & 2025 \\
thermal expansion coefficient \cite{Billone1993} & $\alpha$ & 2.28$\times 10^{-5}$  \\
density (kg/m$^3$) & $\rho$ & 2270 \\
pebble size (mm)* & $d$ & 0.25-0.65 \\
pebble-pebble friction & $\mu$ & 0.2 \\
pebble-wall friction & $\mu_w$ & 0.2 \\
\hline
\end{tabular}
\flushleft
* pebble size distributions used in simulations include (1) mono-sized samples with $0.4\pm0.02$ mm; (2) polydisperse samples of pebbles between 0.25 and 0.65 mm.
\end{table}

\begin{table}[h]
\caption{Interstitial gas properties and material properties of the container wall, at 500 $^\circ$C.}
\footnotesize
\centering
\begin{tabular}{l c l}
\hline \hline
Property & Symbol & Value \\
\hline
purge gas (helium): \\
thermal conductivity (W/mK) \cite{Petersen1970} & $k_g$ & 0.3 \\
gas pressure (kPa) & $p_g$ & 400 \\
\hline
wall (EUROFER \cite{Fokkens2003}): & & \\ 
Young's modulus (GPa) & $E_w$ & 175 \\
Poisson's ratio & $\nu_w$ & 0.3 \\
thermal conductivity (W/mK) & $k_w$ & 29.0 \\
\hline
\end{tabular}
\label{tab:wall}
\end{table}

The height of the simulation cell is 22 mm (along the $x$-axis, adopted from the EU blanket design shown in Fig.\ \ref{fig:DEM}(a)), and its width is 5 mm along the $y$-axis. The depth (the $z$-axis) is varied to accommodate 5,000 pebbles at a given initial packing factor, $\eta$, but the depth dimension is more than 5 times larger than the mean diameter of pebbles. For $y$ and $z$ directions, periodic boundary conditions are applied, while at the $x$ direction elastic wall conditions have been implemented with properties of EUROFER listed in Table \ref{tab:wall}. The wall has a constant temperature of 500 $^\circ$C during the neutron heating stage. Prior to the neutron heating, a uniform heating stage is performed to raise the system temperature from 20 $^\circ$C to 500 $^\circ$C for all the pebbles and walls simultaneously. Then, the neutron heating stage starts with a constant heating profile $\Psi^*$.

Two kinds of size distributions have been considered inside the simulation cells: (1) mono-sized samples with diameters of 0.4$\pm$0.02 mm; (2) polydisperse samples with diameters ranging from 0.25 to 0.65 mm, with a normal distribution with respect to pebble volume. Samples with various initial packing factors have been generated at 20 $^\circ$C with a negligible level of compressive stress, i.e., $p=-\sigma_{ii}/3 < 0.1$ MPa.

To transfer the physical properties of materials listed in Table \ref{tab:mat} and Table \ref{tab:wall} into DEM simulations, some simple scaling laws have been applied to speed up the calculation. Three time scales are considered here: (1) collision time, $t_c=\sqrt{\rho d^2/E}$, describes the typical time for collision events between two elastic pebbles, with elastic modulus of $E$, diameter of $d$ and density of $\rho$; (2) thermal diffusion time, $t_\mathrm{th}= \rho c_p d^2/k_s$, describes the characteristic time to transfer the heat between pebbles, with conductivity of $k_s$ and heat capacity of $c_p$; and (3) heating time, $t_h = \rho c_p \Delta T / \Psi$, provides an estimation for the time to heat a pebble with a temperature increase of $\Delta T$ via a given heating power density $\Psi$. In the current type of problems, the collision time ($t_c$) determines the time step required to achieve meaningful solutions under the explicit scheme of DEM. 

To achieve a larger time step in simulations, a combination of scaled density, size and elastic modulus can be selected by using the scaling laws.
Here, the scaling principle is to ensure the same ratio between the thermal time $t_\mathrm{th}$ and heating time $t_{h}$ for both physical and simulation domains. Therefore, in terms of thermal analysis, the time used in the simulation corresponds to a unit of time in a physical system. Meanwhile, the condition of $t_c \ll t_\mathrm{th}$ and $t_c \ll t_h$ ensures the preservation of  a mechanical quasi-static state during the heat transfer and heat generation. More details can be found in \cite{Gan2012}. The conversion between physical units and simulation units can be found in Table \ref{tab:unit}.

\begin{table}[h]
\caption{Unit system conversion between physical quantities and simulation parameters.}
\label{tab:matsim}
\footnotesize
\centering
\begin{tabular}{l c | l l}
\hline \hline
Property & Symbol & Unit & Simulation\\
\hline
Young's modulus & $E$ & 10$^9$ Pa & 100 \\
force	& $F$ & 10 N & 1 \\
pebble size & $d$ & 10$^{-3}$ m & 1 \\
\hline
conductivity & $k_s$, $k_g$ & 1 W/mK & 1 ** \\
heat capacity & $c_p$ & 1000 J/kg K & 1 \\
density & $\rho$ & 1000 kg/m$^3$ & 1 \\
heating profile & $\Psi$ & 10$^{6}$ W/m$^3$ & 1  ** \\
temperature & $\Delta T$ & 1 $^\circ$C & 1 \\
\hline
\end{tabular}
\label{tab:unit}
\flushleft
** calculated via the scaling law between physical system and simulations: $t_\mathrm{th} /t_\mathrm{th}^\mathrm{sim} = t_h /t_h^\mathrm{sim}$.
\end{table}

\section{III. Results}

In our DEM simulations, samples with two kinds of size distributions of pebbles are considered: (1) mono-sized beds with an initial a packing factor ranging from 62.5\% to 64.5\%; (2) polydisperse beds with packing factor from 63.5\% to 66.0\%. After the initial uniform heating from 20 $^\circ$C to 500 $^\circ$C, three different heating power densities, $\Psi^*$=5, 7, and 8 MW/m$^3$, are applied uniformly across the simulation cell until the cell reaches the steady state. In such cases the variation of the power density along radial direction has been ignored. This can be justified since the ratio between the width and height of the cell is sufficient small, and thus the temperature variation along width is less significant compared with the variation along the cell height. However, under this simplification, the compressive stresses may be overestimated for regions having higher power density due to the restriction of pebble movement along the radial direction to the neighbouring regions having less thermal stresses.

\begin{figure*}[hbt]
\begin{center}
\begin{tabular}{c c}
\includegraphics[width=0.35\textwidth]{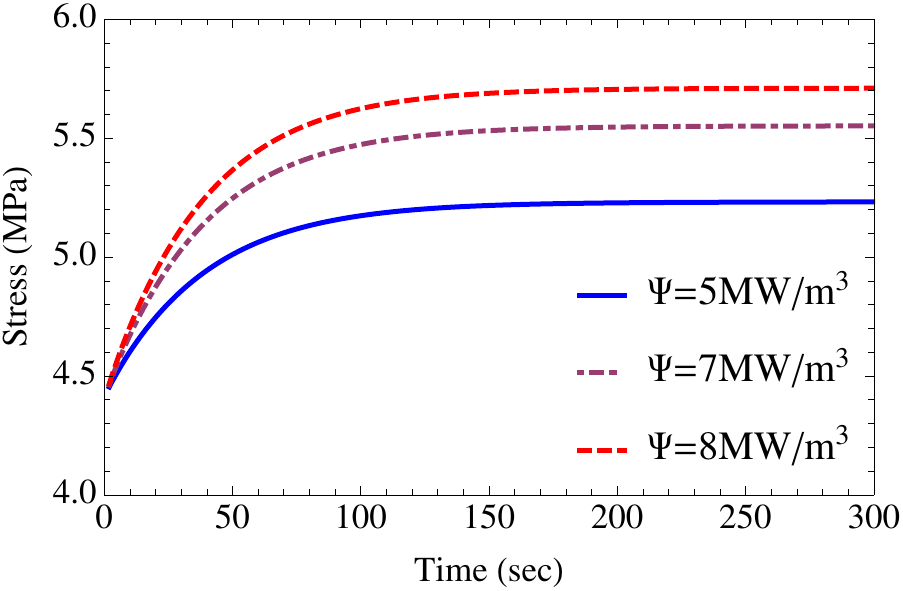} &
\includegraphics[width=0.35\textwidth]{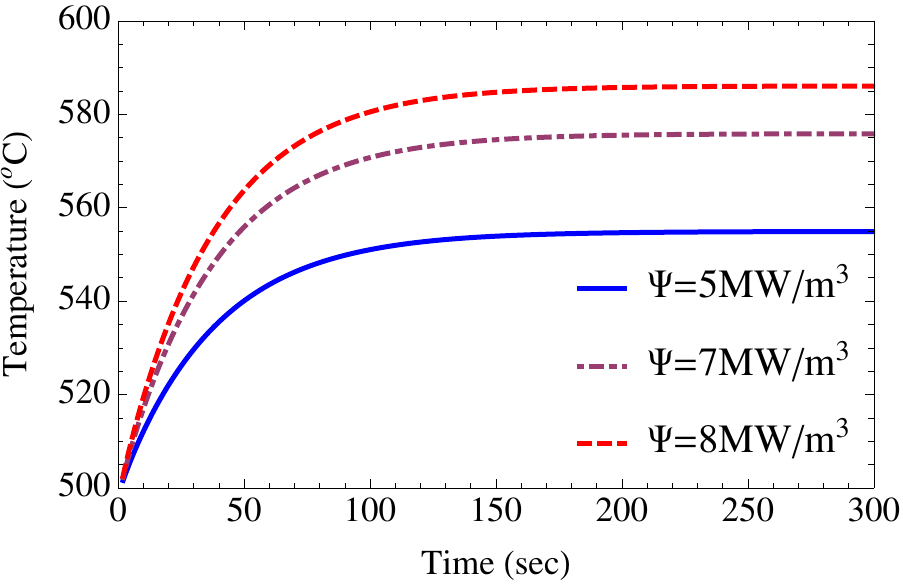}
\end{tabular}
\caption{Evolution of averaged bed temperature (left) and hydrostatic pressure (right) for different heating power densities as a function of neutron heating duration. Sample: mono-sized, $\eta$=63.5\%.}
\label{fig:TSHist}
\end{center}
\end{figure*}

\begin{figure*}[htb]
\begin{center}
\begin{tabular}{c c c}
\includegraphics[width=0.31\textwidth, trim=0 10 0 20]{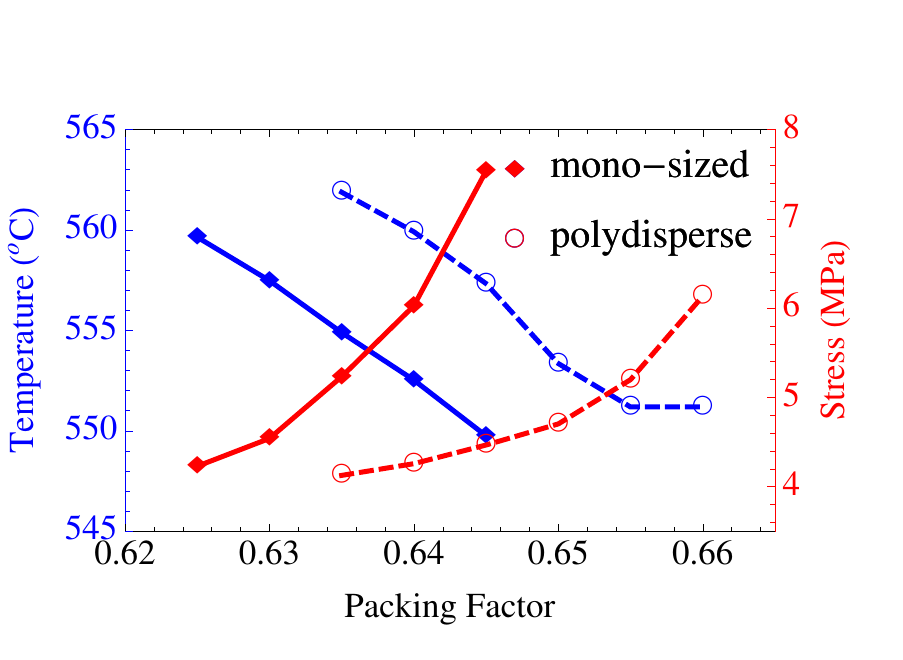} &
\includegraphics[width=0.31\textwidth, trim=0 10 0 20]{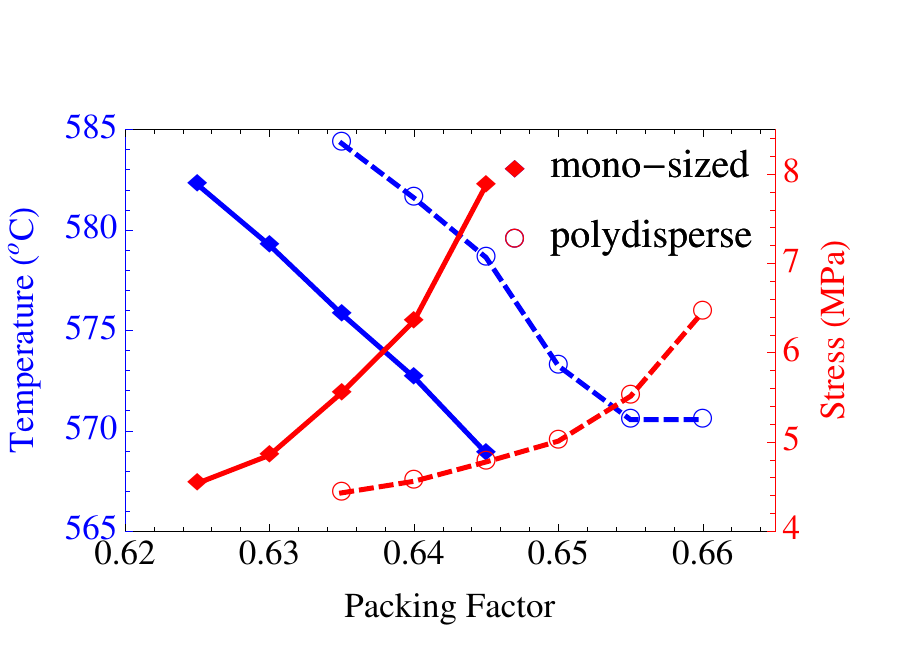} &
\includegraphics[width=0.31\textwidth, trim=0 10 0 20]{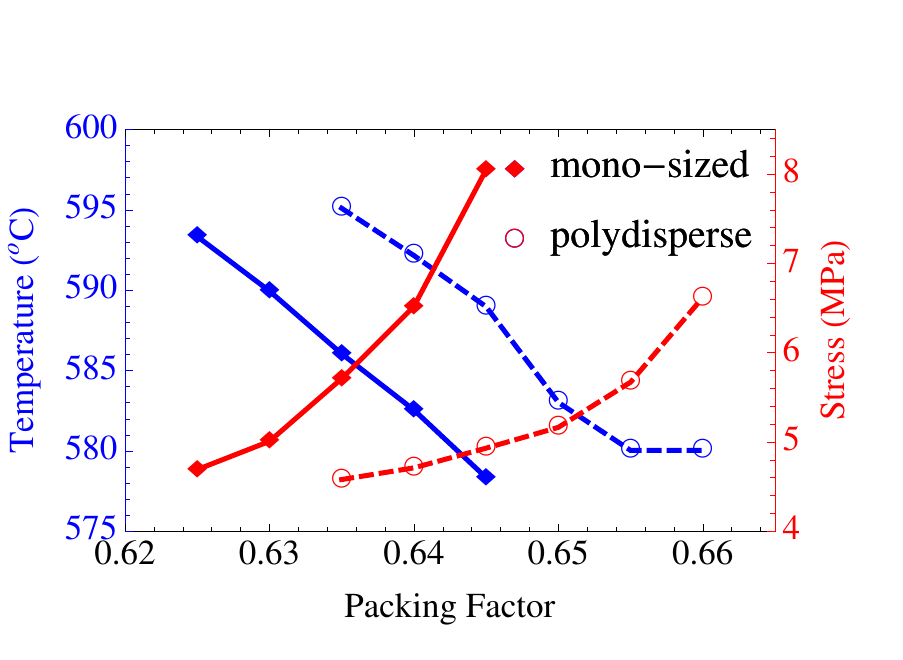} \\
(a) $\Psi^*=5 $ MW/m$^3$ &
(b) $\Psi^*=7 $ MW/m$^3$ &
(c) $\Psi^*=8 $ MW/m$^3$ \\
\end{tabular}
\caption{Thermal stress and average bed temperature, at the steady state, t=300 sec, $\Psi^*$= (a) 5 MW/m$^3$, (b) 7 MW/m$^3$, and (c) 8 MW/m$^3$, as a function of packing factor for mono-sized (filled symbols) and polydisperse systems (empty circles).}
\label{fig:TSChart}
\end{center}
\end{figure*}

Fig.\ \ref{fig:TSHist} shows typical evolutions of the average bed temperature and hydrostatic pressure ($p=-\sigma_{ii}/3$) over operation time from the start-up of neutron heating. The steady states are reached after a few hundred seconds after the start-up. Here, only a mono-sized sample with packing factor $\eta$=63.5\% is shown. Other samples have similar behaviour but with different magnitudes of bed temperature and pressure at the steady state. Prior to the neutron heating stage, the uniform heating from room temperature to 500 $^\circ$C introduces a hydrostatic pressure of 4.5 MPa inside the bed for this mono-sized sample ($\eta$=63.5\%), and the pressure develops to different levels at the steady state,  depending on the corresponding magnitude of the neutron heating power density.
In this analysis, plastic and creep deformations of pebbles have not been considered. The resulting thermal stresses can be reduced if inelastic behaviour is taken into account at the microscopic scale.

Next, the bed temperature and hydrostatic pressure at the steady state (here t=300 sec is selected) for different samples are categorised based on their size distribution of pebbles and plotted against the initial packing factor in Fig.\ \ref{fig:TSChart}. For the same type of size distribution, samples with higher packing factors result in lower bed temperatures but higher stress levels, subjected to a given heating power density. Relative to a mono-sized sample, a polydisperse sample having a similar packing factor can have a lower stress level and a higher bed temperature, under the same neutron heating profile. 
The stress level can be connected to the forces acting on pebbles. Previous studies provide some scaling laws relating to these two parameters \cite{gan2010, Annabattula2012}. 
The correlations between resulting temperature, stress and packing factor provides information for the selection of materials and the design of components for various loading conditions. For instance, the resulting stress state is highly relevant for the determination of the quantitative requirements for crush loads of pebble materials. The average and maximum temperature can be estimated as a key parameter for blanket design and optimisation for tritium breeding and release. In these cases, the grain-scale information provided by DEM proposed here is of great value.

\begin{figure*}[htb]
\begin{center}
\includegraphics[width=0.7\textwidth]{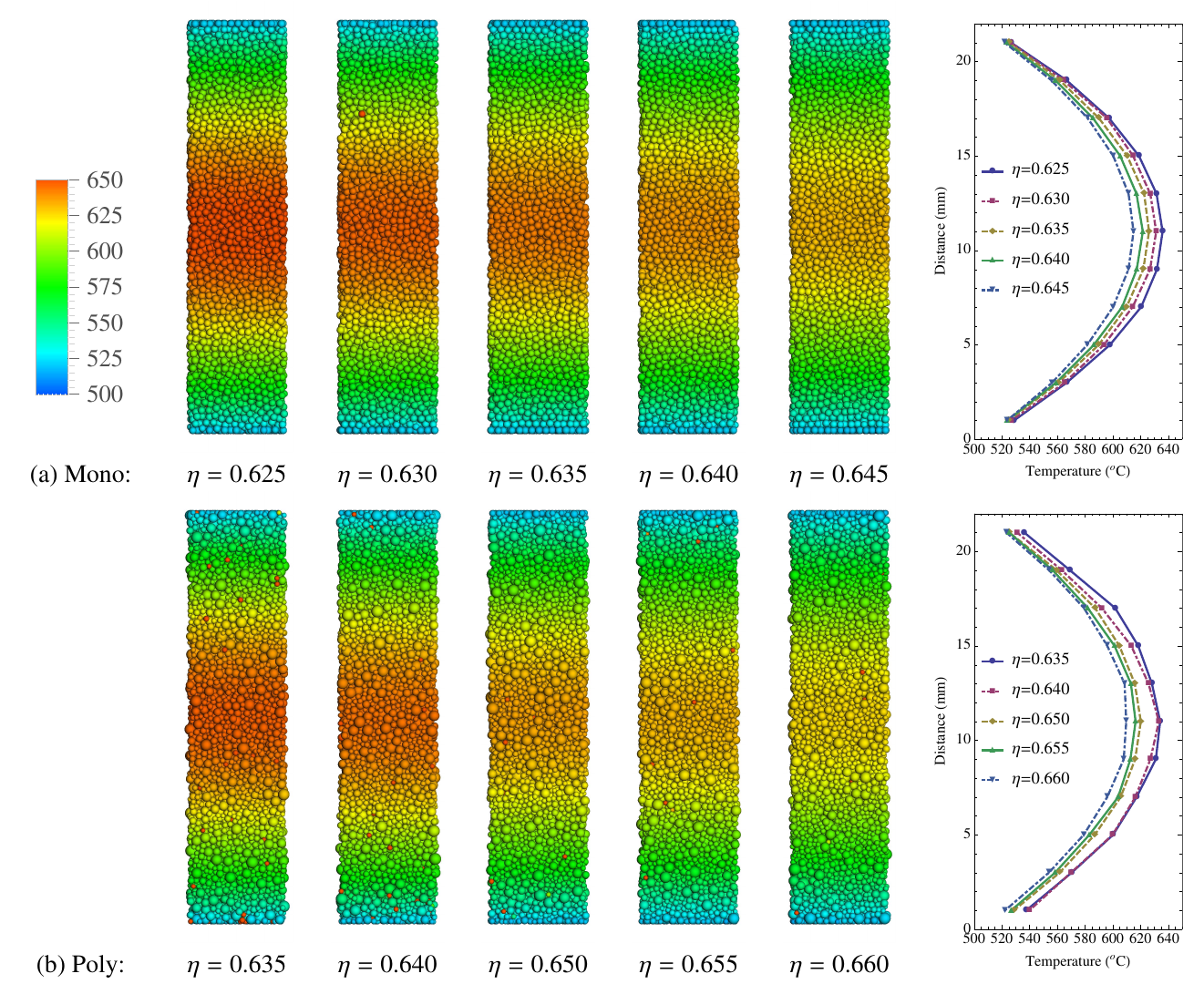}
\caption{Predicted temperature profiles at the steady state (t=300 sec, $\Psi^*$=8 MW/m$^3$): (a) mono-sized samples, (b) polydisperse samples, with different initial packing factors. Pebbles are coloured by their corresponding temperature. In the last column, the volume averaged temperature profiles for each of those samples are plotted against the vertical distance (the walls are located at 0 and 22 mm).}
\label{fig:temp}
\end{center}
\end{figure*}

Fig.\ \ref{fig:temp} shows the temperature distribution at the steady state for different size distributions and initial packing factors. In Fig.\ \ref{fig:temp}, the volume averaged temperature profiles are shown with respect to vertical position, across the ceramic pebble layer. 
With the same initial packing factor, for example $\eta$=0.635 and 0.640, the polydisperse sample presents lower effective conductivity and lower compressive stress level, and thus a higher bed temperature, as compared to the corresponding mono-sized sample, as shown also in Fig.\ \ref{fig:TSChart}(c).
Some "hot spots" are observed in the temperature distributions, and this may result from two factors: (1) a pebble has almost no solid-to-solid contact with its neighbouring pebbles, which leads to a high thermal resistance for heat to be transferred to its surrounding pebbles; (2) the heating power is calculated based on $\Psi^* V_i^\mathrm{cell}$, while for some small pebbles in polydisperse samples, in particular in Fig.\ \ref{fig:temp}(b), the volume occupied by the solid, $V_i^\mathrm{pebble}$, is comparably small in relation to the total Voronoi volume $V_i^\mathrm{cell}$ giving rise to high heating power density in those pebbles. The second factor for the occurrence of "hot spots" can be addressed if the neutronics calculation can consider the heterogeneity within the granular materials, and provide a better way to take into account the transition between the average heating profile and the actual heating power for individual pebbles.
\revision{The maximum temperature found inside the bed is around 650 $^\circ$C, which is lower compared with the finite element (FE) calculations \cite{Hernandez2013}, around 720 $^\circ$C.} The main reasons are: (1) the FE analyses have not considered the effective thermal conductivity with strain dependency which overestimated the maximum temperature; (2) DEM simulations have a pre-heating stage, which results in a hydrostatic pressure of a few MPa and thus a higher bed conductivity even before the neutron heating; and (3) ignoring the region of a large Knudsen Number may result in overestimation of the effective conductivity of the bed.
In general, mono-sized and polydisperse samples have different temperature profiles along the vertical direction, depending on the initial packing factor. For polydisperse samples, temperatures near the wall regions are more sensitive to the initial packing factor. 
\revision{Under the same hydrostatic stress level and heating profile, mono-sized samples have larger temperature gradients, indicating that the bed has a lower effective thermal conductivity, relative to the polydisperse pebble beds.}

\section{IV. Conclusion}
In this study, a thermal discrete element method (Thermal-DEM) has been developed for ceramic breeder materials under neutron irradiation, which includes volumetric heating, thermo-mechanical coupling, and inter-granular thermal conduction. The analysis of ceramic breeder materials subjected to neutron heating under ITER-relevant conditions has demonstrated that the resulting stress state and temperature distribution depend on the type of pebble size distribution of the pebbles and their initial packing factor. At the same heating profile and packing factor, as compared to a mono-sized bed, a polydisperse bed can reduce the stress level, which lowers the risk of crushing pebbles in the bed, but it results in a higher bed temperature. 
\revision{Moreover, this study provides a numerical method to estimate thermo-mechanical responses of beds with various packing factors and pebble size distributions, which can be used as an optimisation tool for the design of solid breeder units.}

\section*{Acknowledgments}
Financial support for this research from the Australian Research Council through Grant No.\ DE130101639 is gratefully appreciated. 
The Thermal-DEM code and scripts for the present study can be downloaded from Github through \url{https://github.com/ganyx/ThermalDEM-TBM}.


\small
\bibliographystyle{ans}
\bibliography{library}

\end{document}